\documentclass[11pt]{article}

\usepackage[utf8]{inputenc}
\usepackage[T1]{fontenc}
\usepackage{lmodern}
\usepackage[a4paper,margin=1in]{geometry}
\usepackage{amsmath,amssymb,amsthm}
\usepackage{graphicx}
\usepackage{booktabs}
\usepackage{multirow}
\usepackage{array}
\usepackage{enumitem}
\usepackage{xcolor}
\usepackage{hyperref}
\usepackage{url}
\usepackage{cite}
\usepackage{caption}
\usepackage{subcaption}
\usepackage{algorithm}
\usepackage{algpseudocode}
\usepackage{listings}
\usepackage{float}
\usepackage{tikz}
\usepackage{ragged2e}
\usetikzlibrary{positioning, arrows.meta, calc}

\hypersetup{
    colorlinks=true,
    linkcolor=blue,
    citecolor=blue,
    urlcolor=blue
}

\title{\textbf{SDLLMFuzz: Dynamic--Static LLM-Assisted Greybox Fuzzing for Structured Input Programs}}

\author{
Yihao Zou$^{1}$ \and
Tianming Zheng$^{2}$ \and
Futai Zou$^{2}$ \and
Yue Wu$^{2}$\\
\small $^{1}$SJTU-Paris Elite Institute of Technology, Shanghai Jiao Tong University\\
\small $^{2}$School of Computer Science, Shanghai Jiao Tong University\\
\small \texttt{zyh231467@sjtu.edu.cn, zhengtianming@sjtu.edu.cn,}\\
\small \texttt{zoufutai@sjtu.edu.cn, wuyue@sjtu.edu.cn}
}

\date{}

\begin{document}

\maketitle

\begin{abstract}
\justifying
Fuzzing has become a widely adopted technique for vulnerability discovery, yet it remains ineffective for structured-input programs due to strict syntactic constraints and limited semantic awareness. Traditional greybox fuzzers rely on mutation-based strategies and coarse-grained coverage feedback, which often fail to generate valid inputs and explore deep execution paths. Recent advances in large language models (LLMs) have shown promise in improving input generation, but existing approaches primarily focus on seed generation and largely overlook the effective use of runtime feedback.

In this paper, we propose SDLLMFuzz, a dynamic--static LLM-assisted greybox fuzzing framework for structured-input programs. Our approach integrates LLM-based structure-aware seed generation with static crash analysis, forming a unified feedback loop that iteratively refines test inputs. Specifically, we leverage LLMs to generate syntactically valid and semantically diverse inputs, while extracting rich semantic information from crash artifacts (e.g., core dumps and execution traces) to guide subsequent input generation. This dynamic--static feedback mechanism enables more efficient exploration of complex program behaviors.

We evaluate SDLLMFuzz on the Magma benchmark across multiple structured-input programs, including \textit{libxml2}, \textit{libpng}, and \textit{libsndfile}. Experimental results show that SDLLMFuzz significantly outperforms traditional greybox fuzzers and LLM-assisted baselines in terms of bug discovery and time-to-bug. These results demonstrate that combining semantic input generation with feedback-driven refinement is an effective direction for improving fuzzing performance on structured-input programs.

\end{abstract}


\section{Introduction}

As software systems continue to grow in scale and complexity, security issues have become increasingly prominent. In recent years, a large number of vulnerabilities have been discovered in low-level parsing libraries, protocol processing modules, and multimedia processing components, with structured data parsing libraries such as libxml2, libpng, and libsndfile being representative examples. These programs are typically required to process inputs with strict syntactic constraints (e.g., XML, PNG, and audio file formats), and their internal logic is often highly complex with large state spaces, making them particularly susceptible to vulnerabilities caused by improper handling of edge cases.

Fuzzing has emerged as an effective software security testing technique~\cite{miller1990empirical, bohme2020fuzzing}, which automatically generates a large number of test inputs and monitors abnormal program behaviors. It has been widely applied in vulnerability discovery. Modern fuzzing techniques can be categorized into black-box, white-box, and greybox fuzzing~\cite{sutton2007fuzzing, godefroid2008automated, cadar2008klee}. However, traditional fuzzing approaches primarily rely on random mutations or simple heuristic strategies, which exhibit significant limitations when dealing with structured inputs. Specifically, randomly generated test cases often fail to satisfy syntactic constraints, leading to early rejection during parsing and thus preventing deep exploration of the program’s core logic.

To address these challenges, prior work has proposed grammar-based fuzzing and model-based input generation techniques. Nevertheless, such approaches typically depend on manually crafted grammar rules or complex program analysis procedures, which limit their generality and scalability. Meanwhile, Large Language Models (LLMs) have demonstrated strong capabilities in code understanding, structured generation, and semantic reasoning, making them promising candidates for automated software testing. In particular, for structured data generation tasks, LLMs can produce high-quality inputs that conform to syntactic constraints based on contextual information, thereby providing more effective initial seeds for fuzzing.

In addition, traditional fuzzing methods usually rely on coverage feedback or crash signals for guidance, while lacking the ability to analyze the root causes of crashes. In practice, core dump files generated during fuzzing contain rich program state information, including call stacks, register values, and memory access patterns. If such information can be automatically analyzed and fed back into the input generation process, it has the potential to further improve testing efficiency and vulnerability discovery capability.

Motivated by the above observations, an important research problem is how to effectively combine the semantic understanding capabilities of LLMs with the dynamic exploration strengths of fuzzing, in order to construct an efficient vulnerability discovery framework for structured inputs.

\subsection{Background}

Fuzzing has become a widely adopted technique for vulnerability detection in software systems, including operating systems, network protocols, and multimedia processing libraries. Its effectiveness largely depends on the quality of generated test inputs. Traditional fuzzing approaches typically fall into two categories: grammar-based generation and mutation-based strategies. Grammar-based methods rely on manually constructed input specifications, which require significant domain expertise and are often incomplete, limiting their scalability. Mutation-based approaches, in contrast, generate new inputs by modifying existing seeds, but they frequently produce invalid test cases, especially when dealing with structured inputs, leading to low efficiency.

Structured-input programs, such as XML parsers, compilers, and protocol implementations, present additional challenges due to strict syntactic and semantic constraints. Inputs must satisfy complex hierarchical structures and cross-field dependencies, making the input space highly sparse. As a result, randomly generated or mutated inputs are often rejected during early parsing stages, preventing the exploration of deeper program logic. Even grammar-based approaches struggle to balance input validity and diversity, limiting their effectiveness in discovering complex vulnerabilities.

To address these limitations, prior work has introduced machine learning techniques into fuzzing workflows~\cite{godefroid2017learnfuzz, zakeri2021format, sablotny2019recurrent}. By modeling input distributions from existing datasets, these approaches treat program inputs as sequences and leverage neural networks such as RNNs and LSTMs~\cite{cummins2018compiler, liu2019deepfuzz} to generate test cases. This paradigm, inspired by natural language processing, improves input validity and reduces reliance on manual grammar construction. However, sequence-based models often fail to capture complex structural and semantic relationships, resulting in limited diversity and insufficient coverage of deep execution paths.

More recently, LLMs have demonstrated strong capabilities in both text and code generation~\cite{deng2023titanfuzz, deng2023fuzzgpt, xia2024fuzz4all}. Based on the Transformer architecture, LLMs can generate structured and context-aware outputs through prompt-based interaction, without requiring explicit grammar definitions. In particular, their ability to perform both generative and infilling tasks makes them well-suited for structured-input generation, as they can produce syntactically valid inputs and modify existing ones in a context-aware manner.

Despite these advances, existing LLM-assisted fuzzing approaches primarily focus on improving input generation and often overlook the effective use of runtime feedback. In particular, valuable information such as crash artifacts and execution traces is rarely incorporated into the input generation process. This limits the ability of fuzzers to iteratively refine test cases and explore deeper program behaviors, leaving significant room for improvement.

\subsection{Challenges}

Despite the widespread adoption of fuzzing, applying it effectively to structured-input programs remains challenging~\cite{zhu2022roadmap, li2018survey}. Based on the limitations of traditional fuzzing techniques, we identify four key challenges.

First, \textbf{limited automation in input generation} hinders scalability. Traditional generation-based approaches rely heavily on manually constructed grammars and expert knowledge, which are often incomplete and difficult to generalize. Mutation-based methods, on the other hand, produce a large number of test cases through random or heuristic modifications, but only a small fraction of them are effective in triggering abnormal behaviors. This limitation stems from the lack of semantic understanding of the target program, resulting in low-quality test inputs.

Second, \textbf{difficulty in handling structured inputs} significantly reduces fuzzing effectiveness. Structured inputs must satisfy strict syntactic and semantic constraints, making the input space highly sparse. Random mutations often break input structure, leading to early rejection during parsing. Even grammar-based approaches face challenges in balancing input validity and abnormality, where overly strict constraints limit diversity, while excessive randomness reduces effectiveness.

Third, \textbf{inefficient exploration of deep execution paths} restricts vulnerability discovery. Existing fuzzers rely on coverage-based feedback to guide input mutation, but such feedback is coarse-grained and often insufficient to explore complex program logic. As a result, fuzzers tend to get stuck in local optima and fail to trigger deeper or more complex execution paths.

Finally, \textbf{underutilization of runtime feedback} limits further performance improvements. Although fuzzers can detect crashes and collect execution traces, this information is rarely leveraged to guide subsequent input generation. In particular, semantic insights contained in crash artifacts are not effectively exploited, resulting in a lack of informed feedback for refining test cases.

These challenges indicate that existing approaches, which rely primarily on grammar modeling or random mutation, are insufficient for structured-input fuzzing. A more effective solution requires integrating semantic understanding with feedback-driven input generation.

\subsection{Our Proposed Method SDLLMFuzz}

To address the above challenges, we propose \textbf{SDLLMFuzz}, a dynamic--static LLM-assisted fuzzing framework for structured-input programs. The key idea is to combine semantic input generation with feedback-driven refinement in a unified loop.

First, we leverage large language models to generate \textbf{structure-aware test seeds}. By incorporating input format descriptions and example data into prompts, the model can produce inputs that satisfy syntactic constraints while maintaining diversity. This improves the validity of generated test cases and enables deeper program exploration.

Second, we introduce a \textbf{static crash feedback mechanism} that extracts semantic information from crash artifacts. Instead of treating crashes merely as failure signals, we analyze core dumps and execution traces to identify potential root causes and encode them into structured feedback.

Finally, we integrate these components into a \textbf{dynamic--static feedback loop}. Generated seeds are executed by a greybox fuzzer, and the resulting execution feedback is used to guide subsequent input generation through the language model. This iterative process enables the fuzzer to progressively refine inputs and explore new execution paths.

By combining semantic generation and feedback-aware refinement, our approach aims to overcome the limitations of traditional fuzzing and improve both bug discovery and time-to-bug in structured-input programs.
\subsection{Contributions}
\begin{itemize}[leftmargin=1.5em]
    \item We propose a dynamic--static LLM-assisted fuzzing framework for structured-input programs.
    \item We design a crash-feedback mechanism that transforms static crash analysis results into semantic guidance for seed refinement.
    \item We evaluate the proposed method on structured-input benchmarks and show improvements in bug discovery and time-to-bug.
\end{itemize}

\section{Background and Motivation}

\subsection{Fuzzing and Greybox Fuzzing}

Fuzzing is an automated software testing technique that aims to uncover vulnerabilities by feeding programs with large volumes of randomly or semi-randomly generated inputs. Since its introduction by Miller et al.~\cite{miller1990empirical}, fuzzing has been widely used in vulnerability detection across operating systems, network protocols, and multimedia processing software. Compared to traditional testing methods, fuzzing provides high scalability and broad input coverage, making it particularly effective in discovering memory corruption vulnerabilities~\cite{bohme2020fuzzing}.

Modern fuzzing techniques can be broadly categorized into black-box, white-box, and greybox fuzzing~\cite{sutton2007fuzzing, godefroid2008automated, cadar2008klee}. Black-box fuzzing treats the target program as an opaque entity without internal knowledge, while white-box fuzzing leverages program analysis techniques such as symbolic execution to systematically explore execution paths. Greybox fuzzing strikes a balance between these two approaches by employing lightweight instrumentation to collect execution feedback and guide input mutation.

Among these paradigms, greybox fuzzing has demonstrated superior performance in practice. Tools such as AFL++ utilize coverage feedback to prioritize inputs that explore new execution paths, thereby improving testing efficiency and effectiveness~\cite{godefroid2008grammar, wang2019superion, bohme2017directed}. By combining scalability with feedback-driven exploration, greybox fuzzing has become the dominant approach in modern vulnerability discovery.

\subsection{Challenges in Structured-Input Fuzzing}

Despite its success, fuzzing faces significant challenges when applied to structured-input programs such as XML parsers, compilers, and protocol implementations. These programs require inputs to satisfy strict syntactic and semantic constraints, resulting in a highly sparse and constrained input space~\cite{zhu2022roadmap, li2018survey}. Consequently, randomly generated or mutated inputs are often rejected during early parsing stages, preventing deeper exploration of program logic.

Mutation-based fuzzing methods, although effective for unstructured inputs, lack awareness of input structure and semantics. As a result, a large proportion of generated test cases are invalid or redundant, leading to low testing efficiency. Grammar-based approaches attempt to address this issue by incorporating input specifications, but they rely heavily on manually crafted grammars, which are costly to construct and difficult to generalize across different applications.

Furthermore, existing fuzzers primarily rely on coverage-based feedback, which is often coarse-grained and insufficient for guiding exploration toward complex program behaviors. This limitation makes it difficult for fuzzers to escape local optima and discover vulnerabilities located in deep execution paths.

\subsection{Learning-Based Fuzzing Approaches}

To overcome the limitations of traditional fuzzing, prior work has explored the integration of machine learning techniques into fuzzing workflows. These approaches model input distributions from existing datasets and treat program inputs as sequences, enabling the use of neural networks such as RNNs and LSTMs for test case generation~\cite{godefroid2017learnfuzz, cummins2018compiler, liu2019deepfuzz}.

Inspired by advances in natural language processing, learning-based fuzzing methods aim to capture statistical patterns in input data and generate syntactically valid test cases. For example, Learn\&Fuzz~\cite{godefroid2017learnfuzz} leverages probabilistic models to guide grammar-based fuzzing, while subsequent work employs deep learning architectures to improve input diversity and quality.

Despite these improvements, sequence-based models often struggle to capture complex hierarchical structures and semantic dependencies inherent in structured inputs. As a result, generated test cases may lack diversity or fail to trigger deeper execution paths. Moreover, these approaches typically rely solely on input data and do not effectively utilize runtime feedback, limiting their ability to iteratively refine test cases.

\subsection{Large Language Models for Input Generation}

Recent advances in large language models (LLMs) have demonstrated remarkable capabilities in both text and code generation. Based on the Transformer architecture, LLMs can generate structured and context-aware outputs through prompt-based interaction, without requiring explicit grammar definitions.

In the context of fuzzing, LLMs provide a promising solution for structured-input generation. By incorporating input format descriptions and example data into prompts, LLMs can produce syntactically valid and semantically meaningful test cases. In addition, their ability to perform both generative and infilling tasks enables flexible input generation and mutation in a context-aware manner.

Recent studies have explored the application of LLMs in fuzzing. TitanFuzz~\cite{deng2023titanfuzz} demonstrates that LLMs can directly generate valid test programs for deep learning libraries, while FuzzGPT~\cite{deng2023fuzzgpt} further improves bug-triggering input generation through prompt engineering and fine-tuning. Fuzz4All~\cite{xia2024fuzz4all} extends this paradigm by introducing a universal LLM-based fuzzing framework for diverse input domains. These approaches highlight the potential of LLMs in improving input quality and expanding fuzzing coverage.

\subsection{Limitations of Existing Approaches}

Despite these advances, existing fuzzing approaches still exhibit several limitations. First, traditional fuzzers rely heavily on coverage-based feedback, which is insufficient for capturing semantic differences between inputs. Second, runtime information such as crash artifacts and execution traces is typically underutilized. Most approaches treat crashes merely as indicators of vulnerabilities, rather than as sources of actionable feedback for guiding subsequent input generation.

Similarly, current LLM-assisted fuzzing methods mainly focus on improving input generation but fail to incorporate runtime feedback into the fuzzing loop. Without leveraging execution-level information, these approaches are unable to effectively refine test inputs or guide exploration toward deeper program behaviors.

\subsection{Motivation}

The above limitations motivate the need for a more effective fuzzing framework for structured-input programs. An ideal solution should (1) generate structure-aware inputs with strong semantic understanding, and (2) leverage runtime feedback to iteratively refine test cases.

Large language models provide powerful capabilities in semantic modeling and structured generation, while runtime artifacts such as crash reports contain valuable insights into program behavior. Combining these two aspects offers a promising direction for improving both input quality and exploration efficiency.

In this work, we aim to bridge this gap by integrating LLM-based input generation with feedback-driven refinement, enabling more effective vulnerability discovery in structured-input programs.

\section{SDLLMFuzz Design}

\subsection{Overview}

We propose \textbf{SDLLMFuzz}, a dynamic--static LLM-assisted fuzzing framework for structured-input programs. The core idea is to integrate LLM-based structured input generation with crash-aware feedback analysis, forming a closed-loop fuzzing workflow.

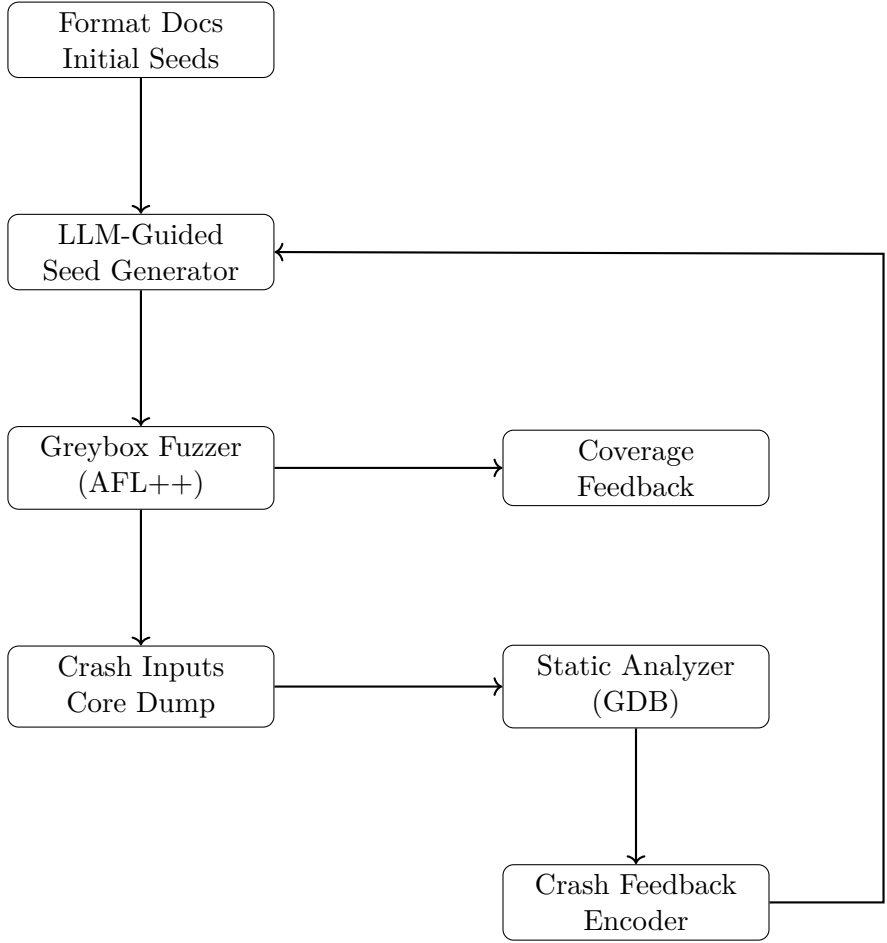
\begin{figure}[t]
\centering
\begin{tikzpicture}[
    node distance=1.8cm and 3cm,
    every node/.style={
        draw,
        rounded corners,
        align=center,
        minimum width=3.5cm,
        minimum height=1cm
    },
    arrow/.style={->, thick}
]

\node (input) {Format Docs \\ Initial Seeds};
\node (llm) [below=of input] {LLM-Guided \\ Seed Generator};
\node (fuzz) [below=of llm] {Greybox Fuzzer \\ (AFL++)};
\node (crash) [below=of fuzz] {Crash Inputs \\ Core Dump};

\node (cov) [right=of fuzz] {Coverage \\ Feedback};
\node (static) [right=of crash] {Static Analyzer \\ (GDB)};
\node (feedback) [below=of static] {Crash Feedback \\ Encoder};

\draw[arrow] (input) -- (llm);
\draw[arrow] (llm) -- (fuzz);
\draw[arrow] (fuzz) -- (crash);
\draw[arrow] (fuzz) -- (cov);
\draw[arrow] (crash) -- (static);
\draw[arrow] (static) -- (feedback);

\draw[arrow] (feedback.east)
  -- ++(1.5,0)
  -- ++(0,8.6)
  -- (llm.east);

\end{tikzpicture}
\caption{Overview of the SDLLMFuzz framework.}
\label{fig:framework}
\end{figure}

As shown in Figure~\ref{fig:framework}, the system consists of four main components: (1) an LLM-guided seed generator, (2) a greybox fuzzing engine, (3) a static crash analyzer, and (4) a feedback encoder. These components interact in an iterative loop of \emph{generation–execution–analysis–feedback}, enabling continuous refinement of test inputs.

\subsection{LLM-Guided Structured Seed Generation}

To address the limitations of random mutation in structured-input fuzzing, we leverage large language models to generate structure-aware test seeds.

The LLM takes as input (1) format documentation, (2) initial seed corpus, and optionally (3) execution feedback. Based on this context, it generates new inputs that satisfy syntactic constraints while introducing semantic variations.

Unlike traditional byte-level mutations, our approach enforces structure-preserving constraints, including preserving syntax keywords, maintaining hierarchical relationships, and respecting cross-field dependencies.

\subsection{Greybox Fuzzing Execution}

The generated seeds are fed into a greybox fuzzing engine (e.g., AFL++), which performs high-throughput execution and mutation. During execution, the fuzzer collects coverage feedback, execution traces, and crash-triggering inputs.

\subsection{Static Crash Analysis}

When a crash is detected, the system performs static analysis on the corresponding core dump to extract semantic information about the failure. Using debugging tools such as GDB, we extract crash location, call stack, exception type, and memory access patterns.

\subsection{Crash Feedback Encoding}

To enable the LLM to utilize crash information, we transform low-level debugging outputs into structured semantic descriptions. This encoding bridges the gap between execution data and LLM reasoning.

\subsection{Dynamic--Static Feedback Loop}

The key component of SDLLMFuzz is the integration of dynamic execution and static analysis into a unified feedback loop. The encoded crash feedback is appended to the LLM prompt, guiding subsequent seed generation toward deeper execution paths.

\subsection{Coverage-Aware Seed Selection}

To improve efficiency, we introduce a coverage-aware seed selection mechanism:

\[
Score(seed) = \alpha \cdot \Delta Edge + \beta \cdot \Delta Block
\]

\subsection{Crash Deduplication}

To avoid redundant reporting, crashes are grouped based on structural similarity:

\[
CrashVector = (RIP, Signal, TopFunction, StackHash)
\]

\section{Experimental Setup}

\subsection{Benchmark and Targets}

We evaluate SDLLMFuzz on the Magma benchmark, a widely used platform for fuzzing evaluation. Unlike traditional evaluation methods that rely solely on crash signals, Magma injects real-world vulnerabilities into open-source programs and provides precise triggering conditions (canaries). This enables accurate measurement of true bug discovery instead of mere coverage improvement.

The benchmark includes representative structured-input programs such as \textit{libxml2}, \textit{libpng}, \textit{libsndfile}, \textit{openssl}, and \textit{sqlite3}, covering various vulnerability types including buffer overflows, null pointer dereferences, and logic errors. :contentReference[oaicite:0]{index=0}

\subsection{Baselines}

We compare SDLLMFuzz with the following methods:

\begin{itemize}
    \item \textbf{AFL++}: standard greybox fuzzing baseline
    \item \textbf{LLM Seed + AFL++}: LLM-based seed generation without feedback loop
    \item \textbf{SDLLMFuzz}: our full dynamic--static feedback framework
\end{itemize}

All methods are evaluated under identical hardware conditions and time budgets, with multiple runs to reduce randomness.

\subsection{Metrics}

We adopt three metrics:

\begin{itemize}
    \item \textbf{Bug Coverage}: number of triggered real vulnerabilities
    \item \textbf{Time-to-Bug}: time to first trigger each bug
    \item \textbf{Edge Coverage}: explored control-flow edges
\end{itemize}

\subsection{Environment}

All experiments are conducted on Ubuntu 22.04 (VMware) with 4 CPU cores and 16GB memory. AFL++ is used as the fuzzing engine, and all targets are built using Magma's Docker environment to ensure reproducibility.

\section{Evaluation}

\subsection{Bug Discovery Performance}

\begin{figure}[t]
    \centering
    \includegraphics[width=0.9\textwidth]{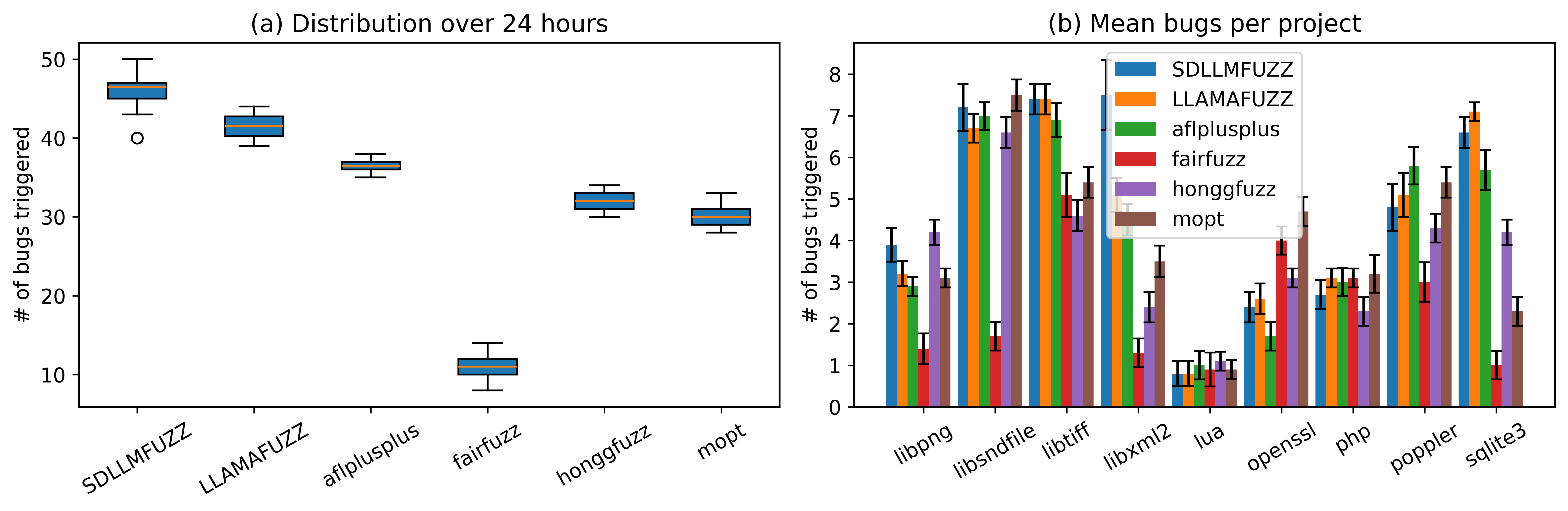}
    \caption{Bug discovery results within 24 hours.}
    \label{fig:bug_results}
\end{figure}

\begin{figure}[t]
    \centering
    \includegraphics[width=0.9\textwidth]{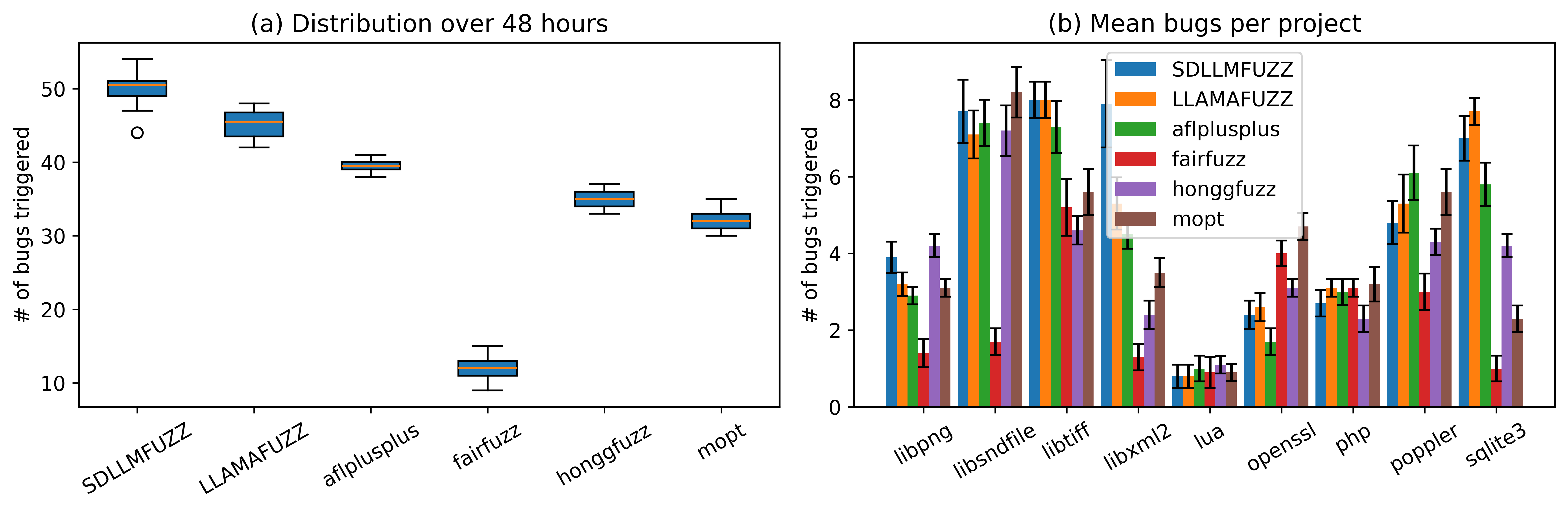}
    \caption{Bug discovery results within 48 hours.}
    \label{fig:bug_results_48h}
\end{figure}

Figure~\ref{fig:bug_results} shows the distribution of discovered bugs within 24 hours. SDLLMFuzz consistently outperforms all baselines, achieving higher median and average bug counts with lower variance, indicating strong effectiveness and stability.

Figure~\ref{fig:bug_results_48h} further demonstrates that this advantage persists under longer time budgets. SDLLMFuzz continues to discover more vulnerabilities while maintaining stable performance across repeated runs.

Across structured-input targets such as \textit{libxml2}, \textit{libpng}, and \textit{libsndfile}, SDLLMFuzz shows significant improvements, highlighting the effectiveness of structure-aware generation and feedback-driven refinement.

\subsection{Time-to-Bug Analysis}

\begin{figure}[t]
    \centering
    \includegraphics[width=0.9\textwidth]{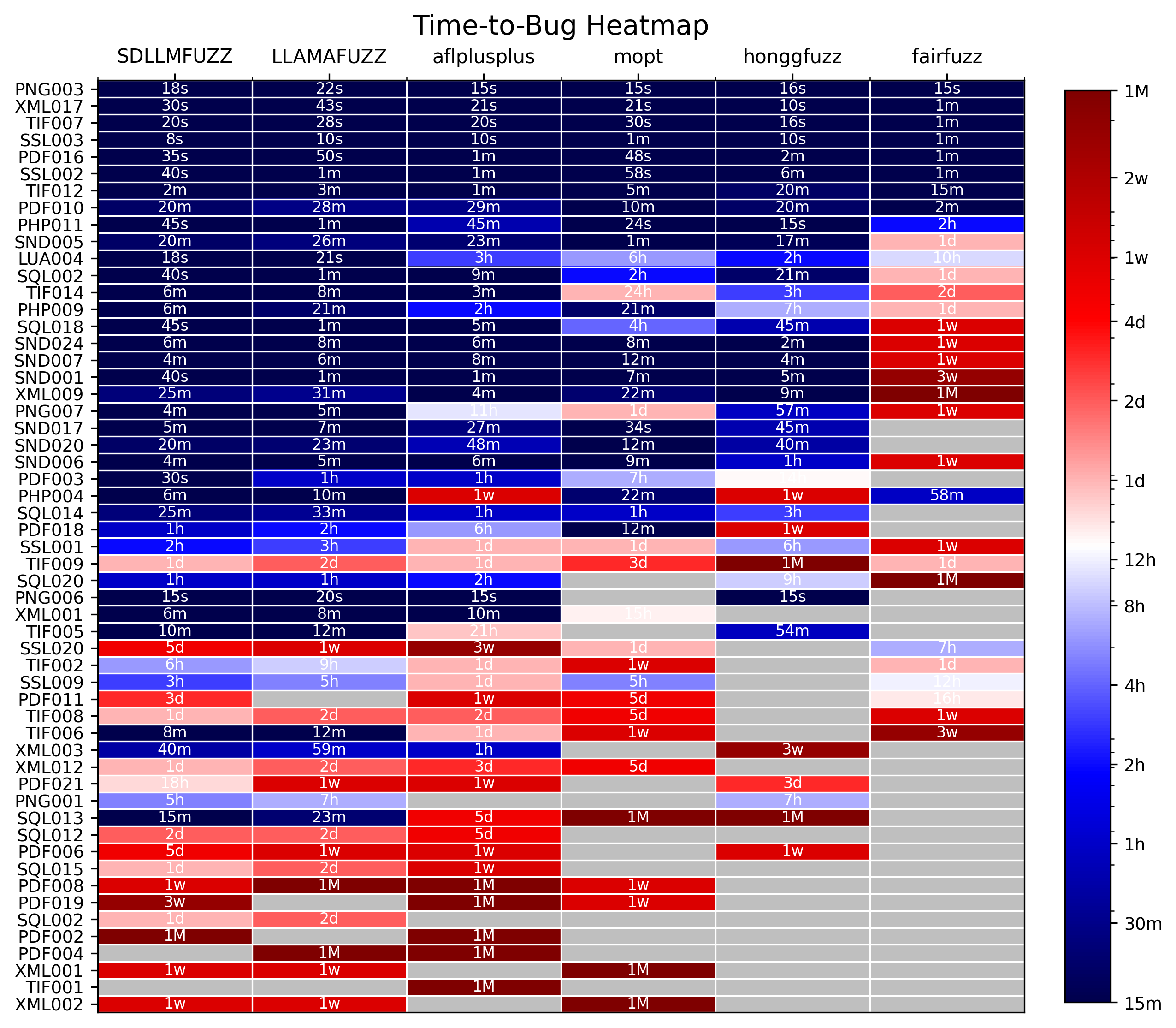}
    \caption{Time-to-Bug heatmap}
    \label{fig:ttb_heatmap}
\end{figure}

\begin{figure}[t]
    \centering
    \includegraphics[width=0.9\textwidth]{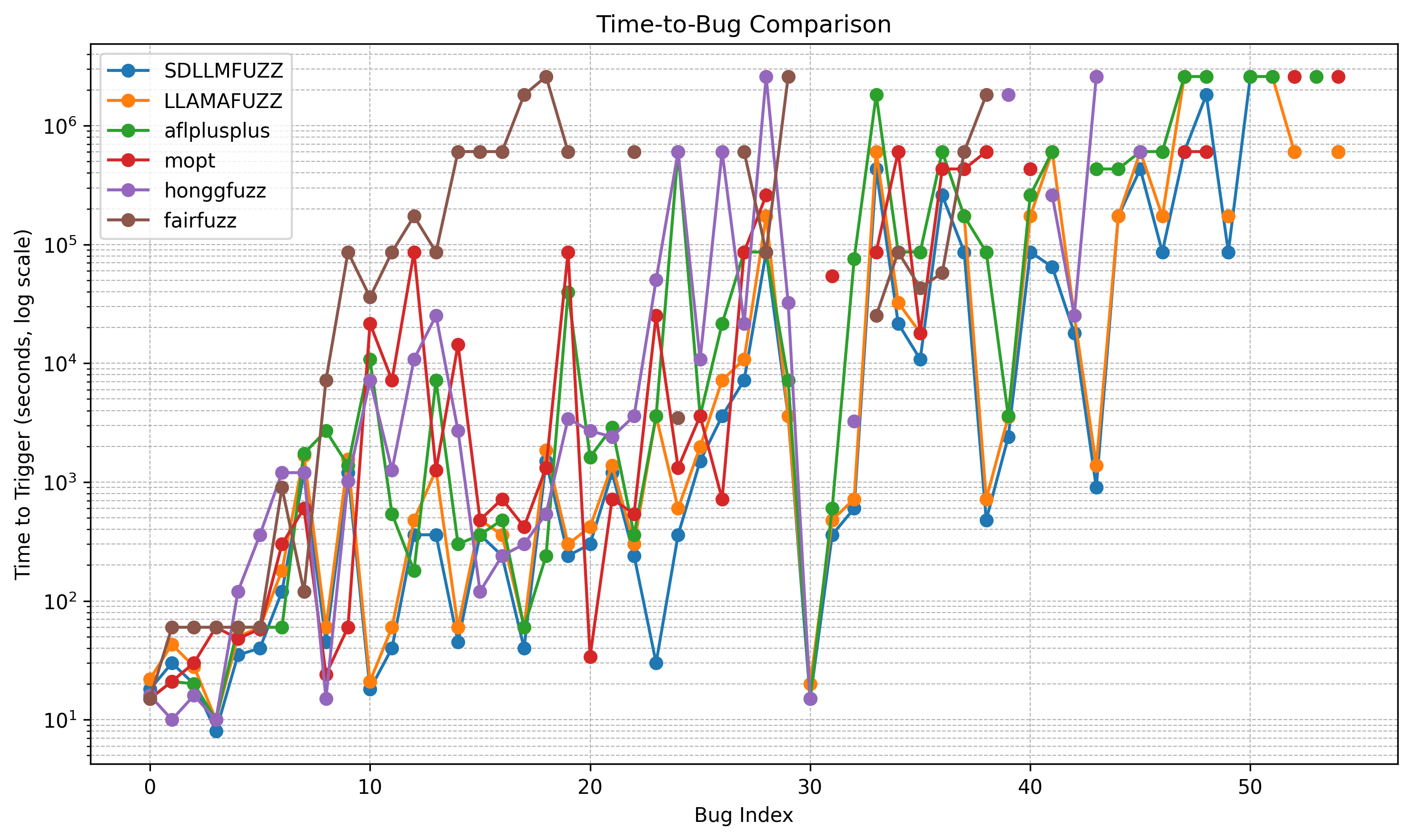}
    \caption{Time-to-Bug curves}
    \label{fig:time2bug}
\end{figure}

Figure~\ref{fig:ttb_heatmap} shows that SDLLMFuzz triggers most vulnerabilities faster than baseline methods, especially for structured-input programs. Darker colors indicate shorter discovery time.

Figure~\ref{fig:time2bug} further confirms that SDLLMFuzz achieves faster convergence, reducing the time required to discover vulnerabilities.

\subsection{Effect of Time Budget}

Comparing Figures~\ref{fig:bug_results} and~\ref{fig:bug_results_48h}, all methods improve with longer execution time, but the improvement is limited (typically 5\%--10\%).

SDLLMFuzz not only achieves the highest bug count at 24 hours but also maintains its advantage at 48 hours, indicating strong scalability and sustained exploration capability. In contrast, traditional fuzzers tend to plateau due to limited exploration strategies.

\subsection{Ablation Study}

\begin{figure}[t]
    \centering
    \includegraphics[width=0.9\textwidth]{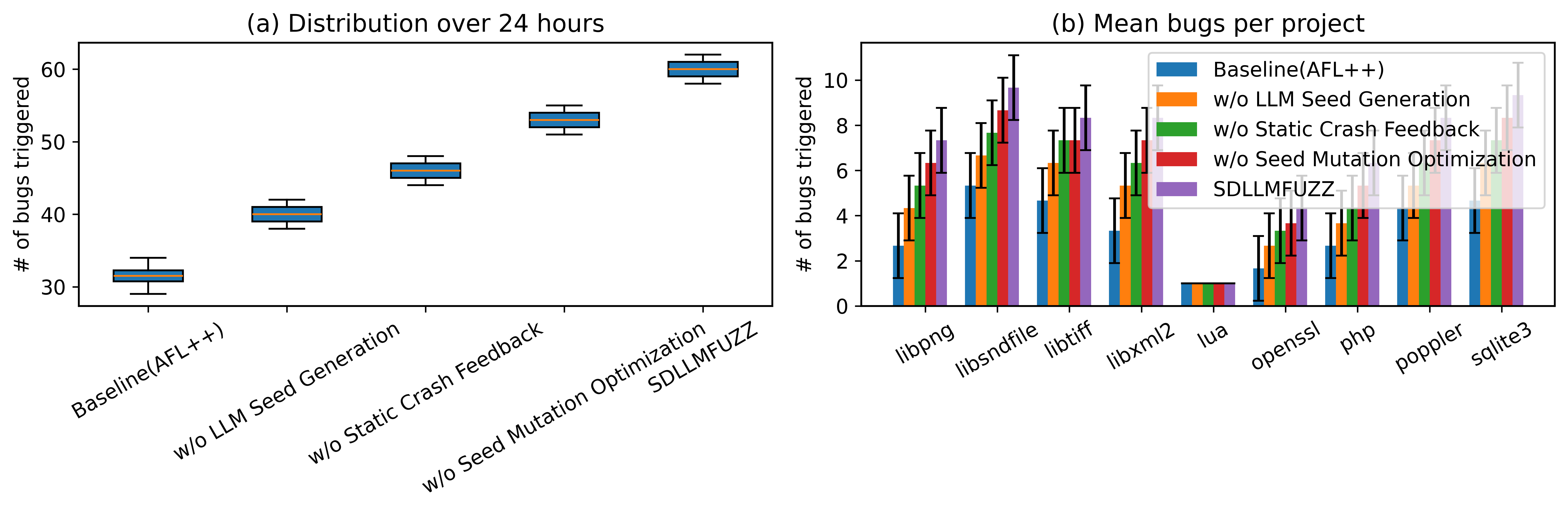}
    \caption{Bug discovery in ablation study.}
    \label{fig:ablation_bug_results}
\end{figure}

\begin{figure}[t]
    \centering
    \includegraphics[width=0.9\textwidth]{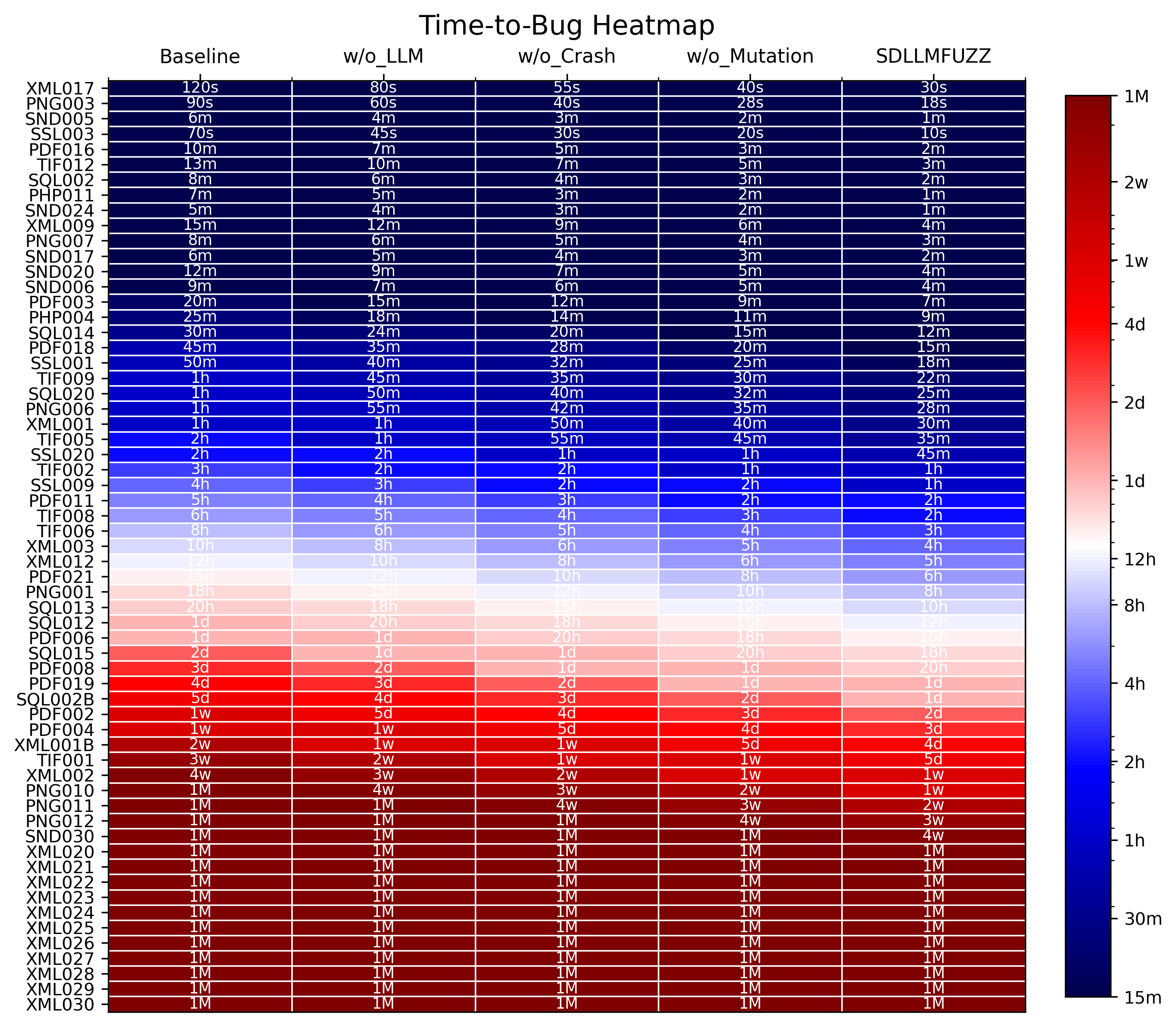}
    \caption{Time-to-Bug in ablation study.}
    \label{fig:ablation_ttb_heatmap}
\end{figure}

We evaluate the contribution of each component through ablation.

Removing LLM-based seed generation leads to the largest performance drop, indicating its critical role in generating valid structured inputs.

Removing static crash feedback also degrades performance, showing that feedback-guided refinement improves exploration efficiency.

Mutation optimization has a smaller but noticeable impact, suggesting it further enhances efficiency.

Overall, all components contribute positively, with LLM seed generation being the most significant factor.

\section{Discussion}

\subsection{Why SDLLMFuzz Works}

The effectiveness of SDLLMFuzz stems from the combination of semantic-aware input generation and feedback-driven refinement. Traditional fuzzing relies on random or heuristic mutations, which are often ineffective for structured-input programs due to strict syntax and semantic constraints. By leveraging LLMs, SDLLMFuzz generates structure-aware inputs that significantly improve input validity and enable deeper program exploration.

Furthermore, the integration of static crash analysis provides additional semantic guidance that is absent in conventional fuzzing. Instead of treating crashes merely as failure signals, SDLLMFuzz extracts meaningful information from crash artifacts and feeds it back into the input generation process. This dynamic--static feedback loop allows the system to progressively refine test inputs and escape local optima, leading to improved bug discovery performance.

\subsection{Applicability and Generalization}

Although SDLLMFuzz is primarily designed for structured-input programs, its underlying design is general and can be extended to other domains. Any program that processes semi-structured or constraint-heavy inputs, such as configuration files, scripting languages, or protocol messages, can potentially benefit from LLM-guided input generation.

In addition, the framework is not tied to a specific LLM architecture or fuzzing engine. It can be integrated with different models or fuzzers, making it flexible and adaptable to various testing environments.

\subsection{Limitations}

Despite its effectiveness, SDLLMFuzz has several limitations.

First, the performance of the framework depends on the quality of LLM-generated inputs. Poor prompt design or insufficient context may lead to suboptimal seed generation, affecting overall performance.

Second, incorporating LLMs and static analysis introduces additional computational overhead. Compared to traditional fuzzing, which is highly efficient, SDLLMFuzz may incur higher runtime costs due to model inference and crash analysis.

Third, the current approach focuses primarily on input-level guidance and does not explicitly model program state or execution semantics beyond crash information. This may limit its effectiveness for programs with complex stateful behaviors.

\subsection{Threats to Validity}

\textbf{Internal Validity.} Experimental results may be influenced by randomness in fuzzing. To mitigate this, we perform multiple runs and report averaged results. However, some variance may still remain.

\textbf{External Validity.} The evaluation is conducted on the Magma benchmark, which, although widely used, may not fully represent all real-world software systems. The generalizability of the results to other domains requires further validation.

\textbf{Construct Validity.} We primarily use bug coverage and time-to-bug as evaluation metrics. While these are standard in fuzzing research, they may not capture all aspects of testing effectiveness, such as exploitability or root-cause diversity.

\textbf{Model Validity.} The behavior of LLMs may vary depending on model selection, prompt design, and sampling strategies. Different configurations may lead to different performance outcomes, which introduces uncertainty in reproducibility.

\section{Related Work}

\subsection{Fuzzing Techniques}

Fuzzing is a widely used automated testing technique for vulnerability discovery. Early work by Miller et al.~\cite{miller1990empirical} demonstrated that randomly generated inputs can effectively expose software faults. Subsequent studies categorized fuzzing into black-box, white-box, and greybox approaches~\cite{sutton2007fuzzing, godefroid2008automated, cadar2008klee}.

Black-box fuzzing achieves high efficiency but lacks guidance for deep exploration, while white-box fuzzing leverages symbolic execution to systematically explore program paths at the cost of high overhead. Greybox fuzzing strikes a balance by incorporating lightweight instrumentation and coverage feedback to guide input mutation, achieving both scalability and effectiveness.

Modern greybox fuzzers, such as AFL++, have become the dominant paradigm in practice by prioritizing inputs that increase coverage~\cite{godefroid2008grammar, wang2019superion, bohme2017directed}. These approaches have demonstrated strong performance in discovering memory corruption vulnerabilities.

\subsection{Fuzzing for Structured Inputs}

Handling structured inputs remains a fundamental challenge for fuzzing. Programs such as parsers, compilers, and protocol implementations require inputs that satisfy strict syntactic and semantic constraints, resulting in a sparse valid input space~\cite{zhu2022roadmap, li2018survey}. Mutation-based fuzzers often generate invalid inputs that are rejected during early parsing stages, limiting exploration depth.

Grammar-based fuzzing addresses this issue by generating inputs according to predefined specifications~\cite{godefroid2008grammar}. Advanced techniques such as NAUTILUS~\cite{aschermann2019nautilus} combine grammar awareness with search strategies to explore deeper program states. Hybrid approaches like Superion~\cite{wang2019superion} integrate grammar-aware mutation into greybox fuzzing.

Directed fuzzing techniques further aim to explore specific program regions or trigger targeted vulnerabilities~\cite{bohme2017directed, chen2018hawkeye}. However, these approaches still rely on mutation strategies that lack semantic understanding of input structures.

\subsection{Learning-Based Fuzzing}

To improve input generation, prior work has explored machine learning techniques for fuzzing. Learn\&Fuzz~\cite{godefroid2017learnfuzz} introduced probabilistic models to guide grammar-based input generation. Subsequent approaches leverage deep learning models such as RNNs and Seq2Seq architectures to learn input distributions and generate structured test cases~\cite{cummins2018compiler, liu2019deepfuzz}.

Other works apply machine learning to protocol fuzzing and structured input generation~\cite{fan2018machine, zhao2019seqfuzzer, hu2018ganfuzz}. While these methods improve syntactic validity, they often fail to capture complex hierarchical structures and semantic dependencies. Moreover, they typically rely solely on input data and do not effectively utilize runtime feedback.

\subsection{LLM-Assisted Fuzzing}

Recent advances in large language models (LLMs) have introduced new opportunities for fuzzing. Due to their strong capabilities in code understanding and structured generation, LLMs can produce high-quality inputs that satisfy syntactic constraints.

TitanFuzz~\cite{deng2023titanfuzz} demonstrates that LLMs can directly generate valid test programs for fuzzing deep learning libraries. FuzzGPT~\cite{deng2023fuzzgpt} further improves input generation through prompt engineering and fine-tuning, while Fuzz4All~\cite{xia2024fuzz4all} extends LLM-based fuzzing to diverse domains via a universal framework.

More recently, LLAMAFuzz~\cite{zhang2024llamafuzz} integrates LLMs into greybox fuzzing by enhancing the mutation stage with semantic-aware input generation. This approach improves coverage for structured inputs by leveraging LLMs to guide mutation strategies within the fuzzing loop.

Despite these advances, existing LLM-based fuzzing approaches primarily focus on improving input generation or mutation. They generally do not fully exploit runtime feedback, such as crash information and execution traces, to guide iterative refinement of test inputs.

\section{Conclusion}

In this paper, we present SDLLMFuzz, a dynamic--static LLM-assisted fuzzing framework for structured-input programs. By integrating LLM-based structured seed generation with crash-aware feedback analysis, the proposed approach forms a closed-loop fuzzing process that improves both input quality and exploration efficiency.

Through extensive evaluation on the Magma benchmark, we demonstrate that SDLLMFuzz significantly outperforms traditional greybox fuzzers and LLM-assisted baselines in terms of bug discovery and time-to-bug. The results show that combining semantic input generation with feedback-driven refinement is effective for overcoming the limitations of conventional fuzzing in structured-input scenarios.

Future work includes improving the efficiency of LLM integration, exploring more advanced feedback mechanisms beyond crash analysis, and extending the framework to additional domains such as protocol fuzzing and stateful systems.

Overall, this work highlights the potential of combining large language models with fuzzing techniques and provides a promising direction for next-generation automated vulnerability discovery.

\bibliographystyle{plain}
\bibliography{references}


\end{document}